\documentclass[11pt,a4paper]{amsart}
\usepackage{amsthm}
\usepackage{amssymb}
\usepackage{graphicx}
\usepackage{enumerate}
\usepackage{marginnote}

\newcommand{\CC}{\mathbb{C}}
\newcommand{\CP}{\mathbb{C}P}
\newcommand{\RR}{\mathbb{R}}
\newcommand{\s}{X_4}  
\newcommand{\DW}{\mathrm{DW}}
\newcommand{\HM}{\mathrm{HM}}
\newcommand{\CY}{\mathrm{CY}}
\newcommand{\mr}{\mathrm}
\newcommand{\ra}{\rightarrow}
\newcommand{\toric}{
\mathcal{T}
}
\newcommand{\OO}{\mathcal{O}}
\newcommand{\DD}{\mathcal{D}}

\usepackage{hyperref}

\theoremstyle{remark}
\newtheorem{remark}{Remark}

\begin{document}

\title[]{
Theory of holomorphic maps of two-dimensional complex manifolds to toric manifolds and type A multi-string theory
}

\author{Olga Chekeres}

\address{University of Connecticut, Department of Mathematics}
\email{olga.chekeres@uconn.edu}

\author{Andrey S. Losev}

\address{Wu Wen-Tsun Key Lab of Mathematics, Chinese Academy of Sciences}
\address{National Research University Higher School of Economics, Moscow, Russia
}
\address{Alikhanov Institute for Theoretical and Experimental Physics (ITEP), Moscow, Russia}
\address{Federal Science Centre ``Science Research Institute of System Analysis at Russian Science Academy'' (GNU FNC NIISI RAN), Moscow, Russia}
\email{
aslosev2@yandex.ru
}

\author{Pavel Mnev}

\address{University of Notre Dame}
\address{St. Petersburg Department of V. A. Steklov Institute of Mathematics of the Russian Academy of Sciences}
\email{pmnev@nd.edu}

\author{Donald R. Youmans}

\address{Albert Einstein Center for Fundamental Physics,
Institute for Theoretical Physics,
University of Bern, Switzerland}
\email{youmans@itp.unibe.ch}

\thanks{The work of A. S. Losev is partially supported by Laboratory of Mirror Symmetry NRU HSE, RF Government grant, ag. 
N\textsuperscript{\underline{o}}  14.641.31.0001.
The work of D. R. Youmans is supported by the NCCR SwissMAP of the Swiss National Science Foundation.}

\date{\today}

\begin{abstract}
We study the field theory localizing to holomorphic maps from a complex manifold of complex dimension 2 to a toric target (a generalization of A model). Fields are realized as maps to $(\mathbb{C}^*)^N$ where one includes special observables supported on (1,1)-dimensional submanifolds to produce maps to the toric compactification. We study the mirror of this model. It turns out to be a free theory interacting with $N_\mr{comp}$ topological strings of type A. Here $N_\mr{comp}$ is the number of compactifying divisors of the toric target. Before the mirror transformation these strings are vortex (actually, holomortex) strings.
\end{abstract}

\maketitle

\section{Introduction}
The goal of this paper is to generalize to complex dimension 2 the theory of holomorphic maps of Riemann surfaces to toric manifolds (an analog of Gromov-Witten or instanton theory).
This may be considered as a first step in the construction of a 4-dimensional quantum field theory of holomorphic maps of complex surfaces to complex
manifolds that we will develop in  an accompanying paper.

Here we would like to consider the case of toric targets and proceed as in \cite{LF}. In this approach we will be able to represent everything in terms of free field theory,
and still find interesting phenomena. In particular, we will see that  the corresponding higher-dimensional theory is a gauge theory  and that the  A-I-B mirror of \cite{LF} is replaced by the multi-string theory of type A.

We will start with a brief reminder of the main constructions of \cite{LF}, then we will show how they are generalized to the case of complex dimension 2, discuss new phenomena
and
point out interesting lines of further development.

\subsection*{Acknowledgements.} We would like to thank Anton Alekseev for fruitful discussions.

\section{Brief review of A-I-B mirror symmetry }
In \cite{LF} Frenkel and Losev considered holomorphic maps from a Riemann surface to a toric variety (we will consider $\CC P^1$ target as an example).
The toric structure on $\CP^1$ means that we consider it as $\CC^*$ compactified by two points (divisors) that we will call $0$ and $\infty$. 
The natural linear structure on $\CC^*  \stackrel{\log}{\simeq}\CC/2\pi i \mathbb{Z}  $ allows one to introduce real coordinates $R$ and $\Phi$ taking values in $\RR$ and $S^1$ respectively. The complex coordinate on the target is
$Z=R+i \Phi$.  
The main idea of \cite{LF} is to consider  each holomorphic map to $\CP^1$ as a holomorphic map from $\Sigma -  \{P_{0,1} , P_{\infty,1}, \ldots ,  P_{0,d} , P_{\infty,d} \}$ to $\CC^*$
where $\Sigma$ is a Riemann surface (we will take $\Sigma$  also to be $\CP^1$ for simplicity), the set $\{ P_{0,1} \ldots P_{0,d} \}$  is the preimage of the divisor $0$ on $\CP^1$ and the set $\{ P_{\infty,1} \ldots P_{\infty,d} \} $ is the preimage of the
divisor $\infty$. In this case the degree of the map is $d$. 
The functional integral contains integration 
over 
the configuration space 
of $2d$  points $\{P_{0,1},\ldots,P_{\infty,d}\}$ in $\Sigma$  (points are allowed to collide).\footnote{\label{footnote 1}
When the genus of the source is zero, the positions of preimages are not restricted, otherwise they are in the kernel of Abel-Jacobi map. In this paper for simplicity we consider simply-connected source.
}
Now we may write the Mathai-Quillen 
 representative for holomorphic maps to $\CC^*$ and add special observables $\OO_0, \OO_\infty$ (called \emph{holomortices}) at preimage points that imitate a
holomorphic map. 
Namely, we will write the action as
\begin{equation}\label{S 2d MQ}
S=-\frac{1}{2\pi}\int_\Sigma p\bar{\partial} Z - \bar{p} \partial \bar{Z} - \pi\bar{\partial} \psi + \bar{\pi} \partial \bar{\psi}
\end{equation}
where $p$ is a $(1,0)$-form and  $\bar{p}$ is a $(0,1)$-form; $p$ and $\bar{p}$ are Lagrange multipliers for holomorphic maps. 
Fermions $\psi$ and $\bar{\psi}$ are the superpartners of coordinates $Z$ and $\bar{Z}$ and fermions 
$\pi$ and $\bar{\pi}$ are the superpartners of Lagrange multipliers.
 Then, one has, schematically,  
\begin{multline}
\delta\left(\begin{array}{c}
\mr{hol\; maps}\; \Sigma\ra \CP^1 \; \mr{of\; degree\;}d
\end{array}\right)\\=
\int_{\Sigma^{2d}\ni (P_{0,1},\ldots, P_{\infty,d})}\int
\mathcal{D}p\,\DD \bar{p}\;\mathcal{D}\pi\,\DD\bar\pi\; e^{-S} \prod_{k=1}^d \OO_0(P_{0,k}) \OO_{\infty}(P_{\infty,k})
\end{multline}

Note that the bosonic part of the action can also be rewritten as 
\begin{equation}
S_\mathrm{bos}= \frac{i}{2\pi} \int_\Sigma -P d \Phi + P * d R   
\end{equation}
where $P=p+\bar{p}$.

Consider some paths $\gamma_i$ connecting points $P_{0,i}$ and $P_{\infty,i}$ and insert into the functional integral the expression
\begin{equation}
\exp  \sum_k  i \int_{\gamma_k} P
\end{equation}

A simple computation shows that in the presence of such an observable one has the  classical solution 
\begin{equation}\label{AIB solution Z=log(...)+C}
Z = \log  \prod_{k=1}^{d} \frac  {(z-P_{0,k})} {(z-P_{\infty,k})} + C
\end{equation}
that is exactly a holomorphic map from  $\CP^1$ to $\CP^1$ of degree $d$.

Now we will describe the mirror map. Integrating over $\Phi$ (we assume for simplicity that the functional integral does not contain observables depending on $\Phi$, i.e., all observables are $U(1)$-invariant), 
we get \begin{equation}
 dP=0
 \end{equation} 
which implies (for $\mathbb{C}P^1$  as a source)
\begin{equation}\label{P=dY}
  P=dY
\end{equation}
which defines a mirror coordinate $Y$   taking values on a circle.\footnote{An observable depending on $\Phi$, for instance $e^{i\Phi}$, becomes a vortex for the field $Y$ in the mirror theory.}
The case of higher-genus source manifolds will be discussed elsewhere.

Observables $\exp  \sum_k   i\int_{\gamma_k} P$ that  were looking non-local become products of local observables
\begin{equation}
\exp \sum_k    i \int_{\gamma_k} P=\prod_{k=1}^{d} \exp\big(-i Y(P_{0,k})\big)  \exp\big(i Y(P_{\infty,k})\big)
\end{equation}
The bosonic part of the action takes the form $\frac{i}{2\pi}\int dY*dR$ and integration over positions of the preimages turns into a deformation of the theory by the superpotential
$\exp(iY)+\exp(-iY)$, see \cite{LF} for further details.

Generalization to general toric manifolds is almost obvious. 
$\CC^*$ is generalized to $(\CC^*)^{N}$ with coordinates $R^a, \Phi^a,\; a=1,\ldots, N$.
Compactifying divisors 
are given by $N$-dimensional integer vectors $\vec{D}_{\beta}
$  (here $\beta$ labels the set of 
compactifying divisors $\{D_\beta\}$);
in the example above 
$N=1$, $D_1=0$ with $\vec{D}_1=+1$ and $D_2=\infty$ with  $\vec{D}_2=-1$. The mirror superpotential is 
$\displaystyle \sum_{\beta}  \exp  \sum_{a=1}^N  D_{\beta}^a Y_a $.

\section{Generalization to complex surfaces}

From the previous section it is clear how to modify the theory for holomorphic maps  $\phi$ of a complex surface $\s$  (subscript $4$ is the real dimension) to a toric manifold.
We will cut out of the surface the preimages of compactifying divisors 
$\phi^{-1} D_{\beta}$ (which are holomorphic curves in $\s$), and get on the complement
the theory of holomorphic maps to $(\CC^*)^N$. 
Below we will consider the case $N=1$; we will 
 present the case of a general toric target in (\ref{eq:general_obs}).

Similarly to (\ref{S 2d MQ}) we may write down the Mathai-Quillen representative
for the delta-form on holomorphic maps inside smooth maps
\begin{equation}
S=-\frac{1}{2\pi}\int_{\s} p\bar{\partial} Z - \bar{p} \partial \bar{Z} - \pi\bar{\partial} \psi + \bar{\pi} \partial \bar{\psi}
\end{equation}
where now $p$ is a (2,1) form on $\s$ and $\bar{p}$ is a (1,2)-form. Similarly, we introduce the 3-form $P=p+\bar{p}$.
 
We have a new phenomenon -- this theory has a gauge symmetry (well-known in the higher-dimensional theory of chiral fields, see 
\cite{CLASH})
\begin{equation}\label{gauge transf}
p \rightarrow p+\bar{\partial}\nu, \quad  
\bar{p} \rightarrow \bar{p}+\partial \bar{\nu}
\end{equation}
where $\nu$ and $\bar{\nu}$ are (2,0) and (0,2) forms on $\s$ respectively, and
there is a similar gauge symmetry for fermions.

The geometrical meaning of such symmetry in the present case comes from the syzygies of holomorphicity equations -- naively, we have twice as many equations as variables,
so naively (for almost complex manifolds) the virtual dimension of the space of almost holomorphic maps is $-\infty$.
However, in the case of integrable complex structures on both source and target (see \cite{paper2} for more on this subject) holomorphicity equations are linearly dependent, and this dependence
 results in syzygies that we see as gauge symmetry for the Lagrange multiplier field. We will discuss the Mathai-Quillen representative for the case of syzygies in \cite{paper2}. 

Due to the linear structure on the target we may fix the gauge symmetry in the standard way -- say, take Lorenz gauge with the help of  a K\"ahler metric on $\s$.

Now, generalizing the 1-dimensional case, we put in the  non-local observable
\begin{equation}
\exp \left( i\int_{\Gamma_3} P \right)
\end{equation}
where $\Gamma_3$ is a 3-manifold such that its boundary is a collection of holomorphic curves in $\s$.
Note that this is required by gauge invariance 
of the observable -- under the gauge transformation (\ref{gauge transf}) it  changes by the factor 
\begin{equation}
\exp\left( i\int_{\partial \Gamma_3} (\nu + \bar{\nu}) \right)
\end{equation}
which is equal to $1$
if the boundary of $\Gamma_3$ is of type (1,1) (that is, the tangent plane to the boundary has type (1,1) in the tangent space to $\s$).

More precisely, $\Gamma_3$ can be a $\mathbb{Z}$-valued 3-chain on $\s$ with boundary $\partial \Gamma_3= \Sigma_{(0)} - \Sigma_{(\infty)}$ where $\Sigma_{(0)}$, $\Sigma_{(\infty)}$ are $(1,1)$-cycles with positive integer coefficients.

Now, it is possible to check that the classical solution $Z$ turns out to be a holomorphic map from the complex surface  $X_4$ to $\CP^1$.

It is possible to construct 
the mirror map. Indeed, integrating out $\Phi$ we get
\begin{equation}
dP=0.
\end{equation}
If $\s$ has no third cohomology, this means that
\begin{equation}\label{P=d Omega}
P=d \Omega
\end{equation} 
where $\Omega$ is a two-form,
so the observable takes the form 
$$\exp\left(-\int_{\Sigma_{(0)}} \Omega\right)  \exp\left(\int_{\Sigma_{(\infty)}} \Omega\right)$$
while
the bosonic part of the action is 
\begin{equation}
\frac{1}{\pi}\int_{\s}   \Omega  \partial \bar{\partial} R.
\end{equation}
From the equations of motion for $R$ we see that $R$ is the real part of a holomorphic function.

Note that here bosonic gauge degrees of freedom correspond to 
 the (2,0)- and (0,2)-form components of $\Omega$ and explicitly decouple from the action.  
Still, we have a new gauge symmetry in bosonic sector
\begin{equation}
\Omega \rightarrow \Omega +d \Lambda
\end{equation}

This completes the story for $N=1$. For general $N$,
the field $\Omega$ couples to the preimages of compactifying divisors as
\begin{equation}
  \prod_\beta
  \exp \left(\int_{\Sigma_{ \beta}}  \sum_{a=1}^{N}  
  D_{\beta}^a \;
  \Omega_a \right) 
\label{eq:general_obs}
\end{equation}
 where $\vec{D}_\beta$ are again seen as vectors in $\mathbb{Z}^N$; 
 $\Sigma_\beta$ is a  $\mathbb{Z}$-valued (1,1)-cycle on $\s$ -- the preimage of $D_\beta$.

  \begin{remark}
  If $H^{1,1}(X_4) \neq 0$, then there exist another global symmetry given by a shift by cohomology:
  \begin{equation}
    \Omega \rightarrow \Omega + \omega^{(1,1)},\qquad \omega^{(1,1)} \in H^{1,1}(X_4).
  \end{equation}
  This symmetry leads to a selection rule: a correlator including an observable as in \eqref{eq:general_obs} vanishes unless 
  \begin{equation}
    \sum_{\beta} 
     \vec{D}_{\beta}\, 
    [\Sigma_{ \beta}] = 0 \quad \in\;\; H_2(X_4)  \otimes \mathbb{Z}^N
  \end{equation} 
\end{remark}

\section{
Bird's-eye view on the emerging theory
}

\noindent\textbf{1.}
We constructed the theory of holomorphic maps from simply-connected $2_\CC$-dimensional complex manifolds to toric targets as a QFT.\bigskip

\noindent\textbf{2.} This provides a strong evidence for existence of the theory of holomorphic maps between complex manifolds in all dimensions of both source and target. We will construct such theory for a  $2_\CC$-dimensional complex source in a forthcoming paper. This theory will be a generalization of 2d $BF$ theory, namely, the theory $\int p\bar\partial X+\mr{c.c.}+\mathrm{fermions}$  with gauge symmetry $p\rightarrow p+\bar{D} \epsilon$.

This generalization  is two-fold: 
\begin{enumerate}[a)]
\item $\int BdA$ is considered as $\int Ad B$ -- as a Mathai-Quillen representative  of the delta-form on 
  constant maps, where gauge symmetries come from syzygies of equations ($\int d\epsilon\, dB =0$). We will also explain the generalization of flat supersymmetric $BF$ theory to general curved targets.\footnote{Abelian $BF$ with curved target, a.k.a. Poisson sigma model with zero Poisson bivector was studied from the viewpoint of formal geometry of the target, \cite{BCM}. It seems that a simpler approach can be used for supersymmetric Poisson sigma model. An analog for such simplification is explained in instantonic theory approach \cite{FLN2}, on the other hand non-supersymmetric curved $\beta\gamma$-system requires formal geometry approach, see \cite{GGW}; see also \cite{Nekrasov2, Movshev,GMS}.}
\item In going to complex dimension $2$,  the exterior derivative $d$ gets replaced by the  Dolbeault operator $\bar\partial$.
\end{enumerate}

Suppose that $\s$ is a Calaby-Yau manofold with Calabi-Yau form $\omega_{\CY}$.
Then we can switch to new variables
\begin{equation}
p=\omega_{\CY} \bar{A},\;\; \pi=\omega_{\CY} \psi_{\DW},\;\; 
\nu=\omega_{\CY} \epsilon_\mathrm{Maxwell},\;\;
H=\omega_{\CY} Z,\;\;  \lambda=\omega_{\CY} \psi
\end{equation}
Then the action takes the form
\begin{equation}
\int H \bar{\partial} \bar{A}  + \lambda \bar{\partial} \psi_{\DW}
\end{equation}
which corresponds to the Mathai-Quillen representative for the Donaldson-Witten abelian theory of holomorphic bundles modulo complex gauge group that is equivalent (on K\"ahler manifolds) to the theory of
self-dual connections modulo the compact group.\footnote{
An observation that holomorphic $BF$ theory with matter with 4-dimensional source leads to holomorphic maps to a toric target was made in \cite{ESW}(Introduction, p.10).
}

However, despite the theories being the same, 
their 
$Q$-differentials are different.
While the 
differential $Q_{HM}$ 
of the theory of holomorphic maps
acts as
\begin{equation}
Q_{\HM} (Z) =\psi, \; \;  Q_{\HM} (\pi)=p
\end{equation}
the differential of Donaldson-Witten theory
acts in the opposite way
\begin{equation}
Q_{\DW} \bar{A}=\psi_{\DW}, \; \;
Q_{\DW} \lambda = H
\end{equation}
because in the theory of holomorphic maps $Z$ is a field and $p$ is a Lagrange multiplier, while in the Donaldson-Witten theory $\bar{A}$ (proportional to $p$) is a field 
and $H$ (proportional to $Z$) is a Lagrange multiplier. 

We find a similar story in supersymmetric Poisson sigma model in real dimension 2. 

Since Donaldson-Witten theory may be generalized to non-abelian case  and the theory of holomorphic maps may be generalized to non-toric targets, we conjecture the existence  of the universal generalized gauge theory that contains Donaldson-Witten and the theory of holomorphic maps as its particular limits. We expect such theory to contain two
differentials -- one that generalizes Donaldson-Witten and another that generalizes the de Rham differential of the holomorphic maps theory.

We expect that the two-dimensional version of such theory is given by the supersymmetric Poisson sigma model that we will study in a separate paper \cite{paper2}.
\bigskip

\noindent\textbf{3.} For toric targets we constructed the analog of  a mirror theory where the gas of points (leading to the superpotential) is replaced by a multi-string theory. 

Namely, holomorphic maps from $\s$ to a toric manifold $\toric$ are expressed by $N_\mr{comp}$ types of topological strings of type A with the target $X_4$. Here $N_\mr{comp}$ is the number of compactifying divisors in $\toric$.

Moreover, consider natural evaluation observables  in the theory of holomorphic maps  $\phi\colon \s\ra\toric$, \;\; $
\int_C \OO^{(\dim C)}_{D_\beta}$,\;\; where $C$ is a cycle in $\s$. 
Namely, consider a map $f\colon \{1,\ldots,k\} \ra \{1,\ldots,N_\mr{comp}\}$, sending $i\mapsto \beta=f(i)$, and consider the correlator
$\displaystyle \left\langle \prod_{i=1}^k \int_{C_i} \OO^{(\dim C_i)}_{D_{f(i)}}
\OO^{(0)}_{p_\toric}(p_{\s})
\right\rangle$ that is the number of maps $\phi$ such that $\phi(C_i)$ intersects $D_{f(i)}$ for all $i$ and such that $\phi (p_{\s})=p_{\toric}$ (here $p_{\s}, p_\toric$ 
are some fixed points in the source and the target; this fixes $(\CC^*)^N$ action on the space of holomorphic maps corresponding to a set of given preimages of compactifying divisors,  cf. the shift by a constant in (\ref{AIB solution Z=log(...)+C})). 

For each $\beta$ consider a type A  correlator in the theory of maps  \mbox{$\Sigma\ra \s$}, $\displaystyle n_{\beta} = \left\langle \prod_{i\in f^{-1}(\beta)} \int_\Sigma \OO_{C_i}^{(2)} \right\rangle$
-- the number of holomorphic maps passing through cycles $C_i$, with $f(i)=\beta$. In $n_\beta$ we are also integrating over complex structures on $\Sigma$ and summing over topological types of $\Sigma$ (possibly disconnected, with connected components of any genus).

The relation between the theory of holomorphic maps $\s\ra \toric$ and maps $\Sigma\ra \s$ is schematically given by 
\begin{equation}
 \left\langle \prod_{i=1}^k \int_{C_i} \OO^{(\dim C_i)}_{D_{f(i)}}
\OO^{(0)}_{p_\toric}(p_{\s})
 \right\rangle=\prod_{\beta=1}^{N_\mr{comp}} n_\beta 
\end{equation}
In some sense, the theory of holomorphic maps $\s\ra \toric$ looks like a second quantized field theory for $N_\mr{comp}$ types of strings. We are planning to study this unexpected  phenomenon in the future.

\includegraphics[scale=0.67]{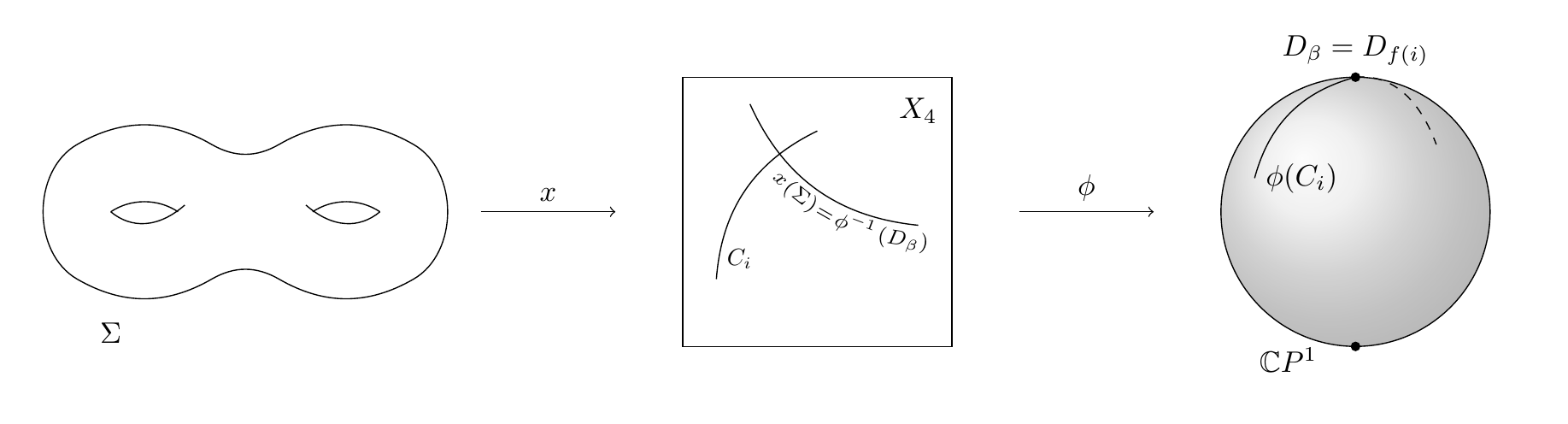}

\bigskip


\noindent \textbf{4.} One is tempted to make the following speculation.
The mirror theory is a theory of several types of strings coupled to free field theory. 
It looks similar to compactifications of M-theory that are, roughly speaking, Lagrangian field theories coupled to a set of extended objects. It is desirable -- but not known -- how to understand compactifications of M-theory as a field theory in Segal's sense. The theory that we consider seems to provide a much simpler example of this phenomenon. One can go even further. Among similar examples modeling M-theory are instantonic strings (instantons exist in codimension $4$) in 6d gauge theories. 
Interestingly, such strings can be considered as 
 fundamental strings, bound to a 6d NS5-brane in type IIB theory (that is why they are called ``little strings'' in \cite{LMS}). More recently people considered vortex strings. Since vortices have codimension 2, they exist in 4d theories and there are strong arguments that they are also bound states of fundamental strings \cite{HT,SY2}. 
Strings that we consider are also vortex strings (in \cite{LF} the corresponding object was called ``holomortex'') and seem to be the topological sector of Shifman-Yung's little strings. Certainly, we are planning to investigate this relation further.\bigskip

\noindent \textbf{5.} It seems very interesting to study tropicalization \cite{Mikhalkin1,Mikhalkin2} of the constructed theory. Not only it allows computations and turns algebraic geometry into combinatorics -- tropicalization of a string theory is also a field theory. Thus, after tropicalization, a 4d theory with strings becomes a 2d theory with particles that may be related to Feynman diagrams in conventional 2d theory, despite not being Lorentz-invariant. We plan to study this tropicalization elsewhere. Tropicalization may lead to an approach to a higher-dimensional analog of WDVV. 
Tropicalization of the theory on a toric manifold is a field theory on its moment polytope.
In particular, for the source of complex dimension $2$, 
we will get a field theory on a convex polygon. Feynman diagrams will be tropical curves that are 
graphs with straight edges. Summing up these graphs will allow to replace tropical strings by ``integrating in''  another field (in the terminology of Seiberg) -- this is an analog of non-perturbative summation of instantons in complex \mbox{dimension 1}. A theory of this type 
was constructed in \cite{LS}; for the case of tropical curves in  toric manifolds, see \cite{LL}.
\bigskip

\noindent \textbf{6.} It is clear from the discussion above that consideration of holomorphic maps is not restricted to source manifolds of complex dimension 2. However, it may be technically more complicated. For instance, in tropicalization 
graphs with straight edges
are replaced by polyhedral complexes. That is why we restrict to complex dimension 2.
\bigskip

\noindent \textbf{7. Disclaimer A.} Actually, we are dealing not with  holomorphic maps but rather with Drinfeld's holomorphic quasi-maps 
(since e.g. for $\CP^1$ target the preimages of $\{0\}$ and $\{\infty\}$ are allowed to intersect).\bigskip

\noindent\textbf{\phantom{7.} Disclaimer B.}  The situation becomes more complicated when the source is not simply-connected. In particular, formulas \eqref{P=dY} and \eqref{P=d Omega} have corrections due to 1st and 3rd cohomology of the source. We will address this issue in the nearest future.\bigskip

\noindent \textbf{8.} Holomorphic quasi-maps between two toric  varieties of any dimension can be effectively described, and an analog of Givental-Nekrasov theory \cite{Givental,Nekrasov} can be constructed. It may lead to another approach to higher dimensional version of WDVV theory.

\end{document}